\documentclass[prb,preprint,superscriptaddress,twocolumn,10pt]{revtex4-1} 

\usepackage{amsmath}  
\usepackage{amsfonts} 
\usepackage{graphicx} 

\begin{document}


\title{The Effect of Charge Discretization on the Electrical Field inside a Conductor}


\author{Nam H. Nguyen}
\affiliation{Duke Univeristy, Durham, NC 27708, USA }

\author{Quy C. Tran} 
\affiliation{High School for Gifted Students, Vietnam National University, Hanoi 100000, Vietnam}

\author{Thach A. Nguyen}
\affiliation{Le Hong Phong High School for the Gifted, Ho Chi Minh 700000, Vietnam}

\author{Trung V. Phan}
\email{trung.phan@yale.edu}
\affiliation{Department of Molecular, Cellular, and Developmental Biology, Yale University, New Haven, CT 06520, USA}

\begin{abstract}
We show how the electrical field inside the conductor changes as a function of the number of charged-particles. We show that the non-vanishing electrical field is concentrated near the surface of the conductor, at a shallow depth on the same order of magnitude as the separation between charges. Our study has illustrated the effect of charge discretization on a fundamental emergent law of electrostatics.
\end{abstract}

\date{\today}


\maketitle 

\section{Introduction}

It is often stated in introductory physics textbooks that the electrical field inside a conductor should be zero\cite{griffiths2005introduction,purcell2013electricity,jackson1999classical,landau2013electrodynamics, maxwell1873treatise}, but that can only be true if the charge is a continuous distribution. Consider a simple classical model (which is used in the famous Thomson problem \cite{thomson1904xxiv}), in which the charge is made of identical point-particles and the conductor is a confined volume in space that the charged-particles can freely move around. If there is only a single charged-particle, no matter where it is, there will always be a non-vanishing electrical field inside the conductor. As the number of charged-particles increases, we approach the continuum limit, thus it can be expected that the electrical field inside the conductor will eventually disappear at electrostatic equilibrium (the state of minimum electrostatic potential energy, in which the charge-particles are concentrated on the surface \cite{levin2003charges}).

\ \ 

It is curious to ask how the electrical field $\vec{E}$ gets suppressed by the number of charged-particles $\mathcal{N}$. This is the main question that we are going to explore in this note. For convenience and simplicity, we will only work with an even number $\mathcal{N} \equiv 2N$ and a symmetric-shaped conductor. Since how exactly charged-particles distribute on the surface of a three-dimensional spherical conductor is still an unsolved problem \cite{smale1998mathematical}, we will first look at the situation in two-dimensional space (2D) instead, in which the conductor is a circular area. Then we will argue about what might happen in three-dimensional space (3D).

\ \ 

\section{Two-Dimensional Space}

\ \ 

\begin{figure}[!htb]
\centering
\includegraphics[width=0.45\textwidth]{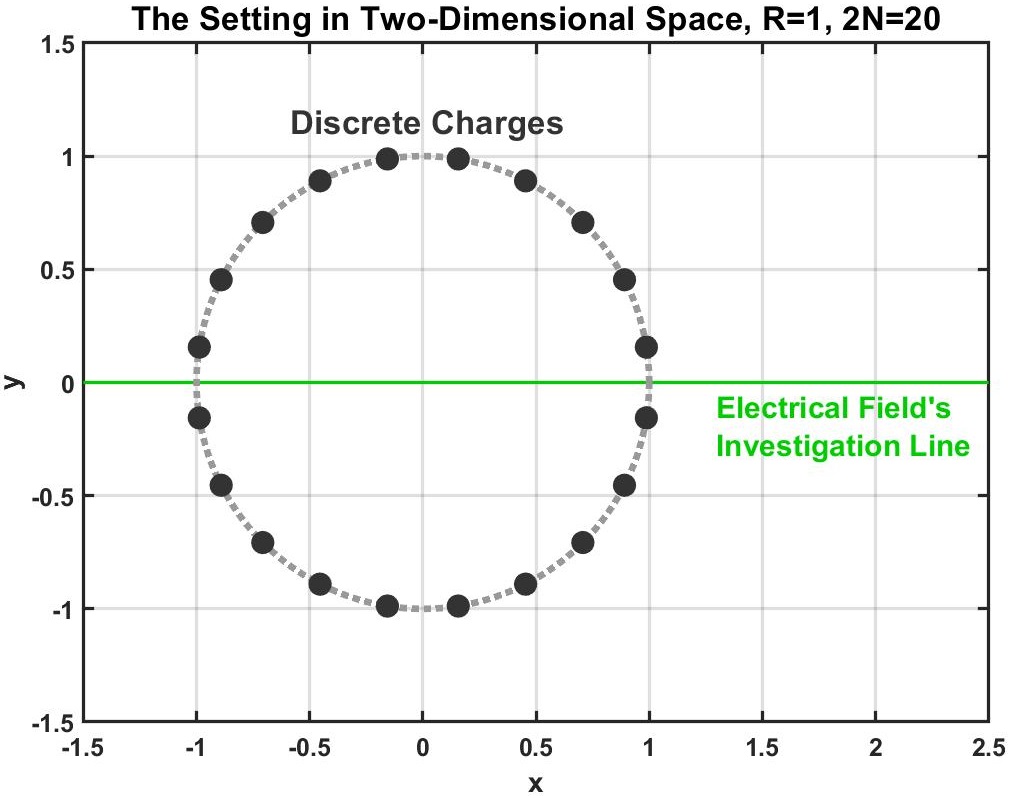}
\caption{The physical setting of interests in two-dimensional space, in which the electrostatic energy is minimum.}
\label{fig:2D_Setting}
\end{figure}

From the Gauss's law \cite{gauss1877theoria}, we can calculate that the electrical field created by a charged-particle of quantized charge $q$ at distance $r$ is given by $|\vec{E}|=q/2\pi \epsilon_0 r$ in 2D. The lowest-energy configuration for charge distribution in a circular area has all charged-particles concentrated on the boundary and evenly separated \cite{amore2019thomson}. We can choose a Cartesian coordinate O$xy$ in which the origin O is located at the center of the conductor and the $x$-axis is a symmetry axis, as described in Fig. \ref{fig:2D_Setting}. The angular position (with respect to the $x$-axis) of the charged-particles are given by $\theta_n = \pi (n-1/2)/N$. For the total charge is $Q$, the individual charge should be $q=Q/2N$. We have the freedom to define the conductor's radius $R=1$ to be the unit-length, the total charge $Q=1$ to be the unit-charge and $1/2\pi \epsilon_0=1$ to be the unit-value of electrical field.

\ \ 

\ \ 

\begin{figure}[!htb]
\centering
\includegraphics[width=0.45\textwidth]{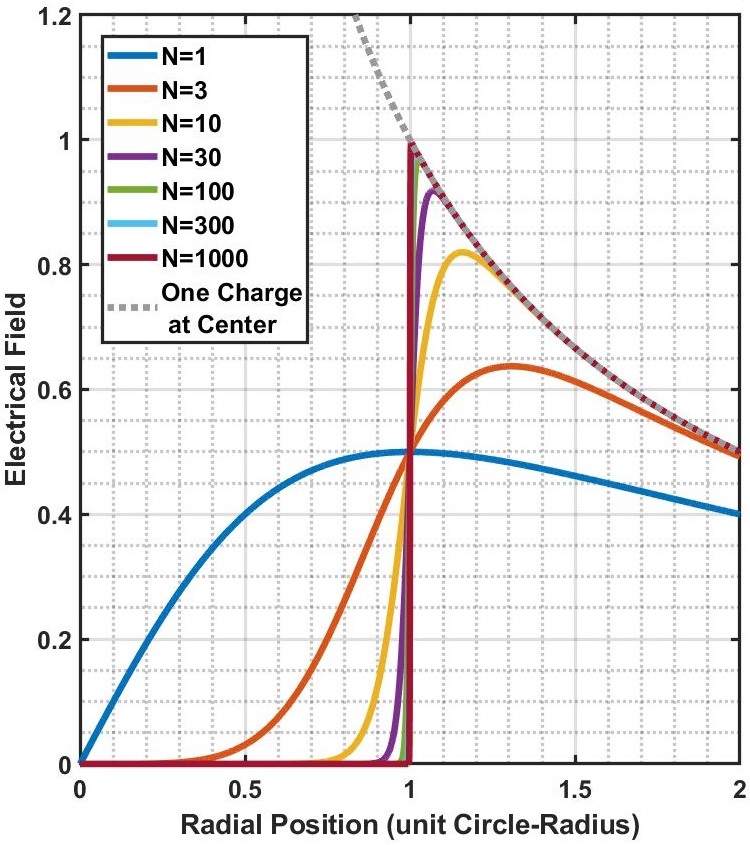}
\caption{The measurements of electrical field depending on how far it is from the center of the conducting circle. As $N$ goes up, the electrical field inside the conductor getting more and more suppressed.}
\label{fig:Electrical_Field_All}
\end{figure}

To understand what is going on inside the conductor, we can probe by investigating the electrical field along the $x$-axis. The electrical field created at position $x \geq 0$ can be calculated from the series summation:
\begin{equation}
E(x) = \frac1N \sum^N_{n=1} \frac{x - \cos\theta_n}{\left(x - \cos\theta_n \right)^2 + \sin^2 \theta_n} \ ,
\label{E_in_x_2D}
\end{equation}
where $\vec{E}=E \hat{x}$ and $\hat{x}$ is a unit-vector pointing in the positive $x$-direction. This function $E(x)$ can be numerically investigated, as shown in Fig. \ref{fig:Electrical_Field_All}. 
\ \ 

\ \  

We can now make some observations. For any value $N>0$, at far away $x\gg 1$ the electrical field asymptotically approach $E(x)\rightarrow 1/x$. For large value $N \gg 1$, the electrical field effectively vanish inside the conductor and only exists near the boundary. The shallow region of non-zero electrical field has a characterisitic depth $\lambda$, which can be estimated from the electrical field $\vec{E}$ and its gradient $\nabla \vec{E}$ at the boundary:
\begin{equation}
\lambda \sim \frac{E(x)}{\partial_x E(x)} \bigg|_{x=1} \ .
\label{lambda_in_x_2D}
\end{equation}
Let's do a series expansion of Eq. (\ref{E_in_x_2D}) at around the boundary $x = 1+z$:
\begin{equation}
\begin{split}
&E(1+z) = \frac1{N} \sum^N_{n=1} \frac{(1+z) - \cos\theta_n}{\big[(1+z) - \cos\theta_n \big]^2 + \sin^2 \theta_n} 
\\
& \ \ \approx \frac1N \sum^N_{n=1} \left[ \frac12 - \frac12\left(1 - \frac1{1-\cos\theta_n} \right) z + \mathcal{O}(z^2) \right] 
\\
& \ \ = \frac1{2} - \left( \frac12 - \frac1{2N} \sum^N_{n=1} \frac1{1-\cos\theta_n} \right) z + \mathcal{O}(z^2) \ .
\end{split}
\label{E_in_z_2D}
\end{equation}
This analytical result means that at the boundary the electrical field is always $E(1)=1/2$ for any value $N>0$ and the electrical field's gradient there depends on the series summation:
\begin{equation}
S_N \equiv \sum^N_{n=1} \frac1{1 - \cos\theta_n} \ .
\label{final_sum_2D}
\end{equation}
It is known that the $N$-th Chebyshev polynomials of the first kind \cite{arfken1985chebyshev}:
\begin{equation}
T_N (\zeta) = 2^{N-1} \prod^N_{n-1} \Big( \zeta - \cos\theta_n \Big) 
\label{Chebyshev}
\end{equation}
has the property $T_N(\cos \xi ) = \cos(N\xi )$. Our summation of interests $S_N$ can then be calculated exactly from taking the limit:
\begin{equation}
\begin{split}
&S_N = \lim_{\zeta \rightarrow 1} \partial_\zeta \ln \Big[ T_N(\zeta) \Big]= -\lim_{\xi \rightarrow 0} \frac{\frac1{\sin\xi}\partial_{\xi} T_N(\cos\xi)}{T_N(\cos \xi)}
\\
& \ \ \ \ \ \ \ \ = -\lim_{\xi \rightarrow 0} \frac1{\sin\xi} \frac{\partial_\xi \cos(N\xi)}{ \cos(N\xi)} = N^2 \ .
\end{split}
\label{exact_final_sum_2D}
\end{equation}
We can apply this results into Eq. (\ref{E_in_z_2D}) and Eq. (\ref{lambda_in_x_2D}) to arrive at the estimation:
\begin{equation}
\lambda \sim \frac{E(1)}{\partial_z E(1+z) \Big|_{z =0}} = \frac{1/2}{(N-1)/2} = \frac1{N-1} \ .
\end{equation}
At large value $N \rightarrow \infty$ we obtain the inverse relationship $\lambda \sim 1/N$. This depth is about the same order of magnitude with the spacing $a \approx \pi/N$ between charged-particles on the boundary. The physical interpretation can be explained as followed: the continuum limit breaks down when one observes the electrical field close enough to boundary of the conductor such that the discretization of charge becomes visible.

\ \ 

\section{Three-Dimensional Space}

\ \ 

\begin{figure}[!htb]
\centering
\includegraphics[width=0.45\textwidth]{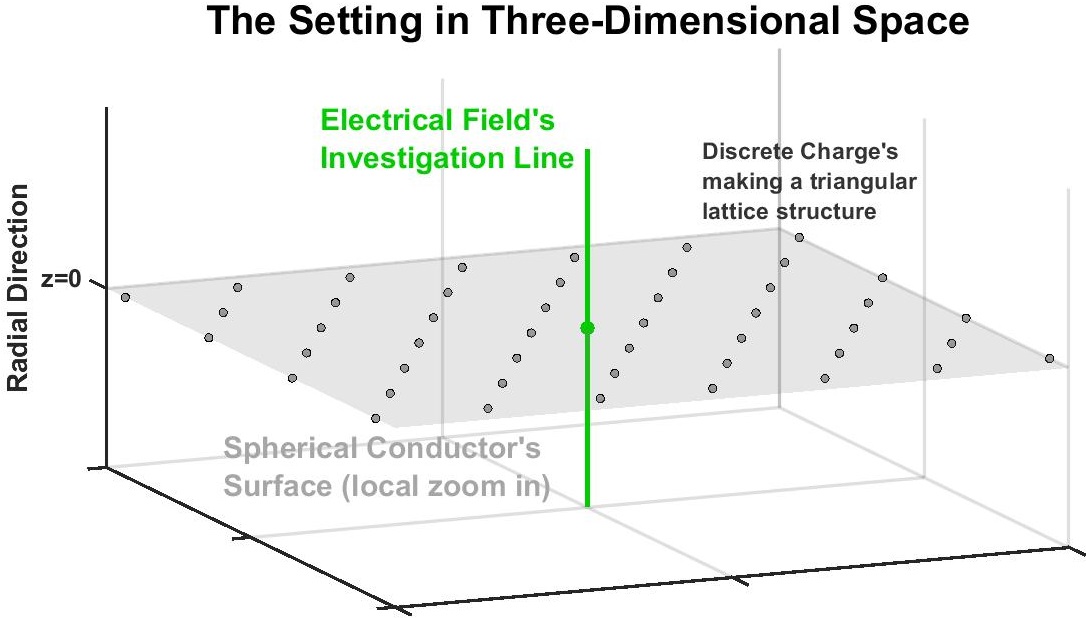}
\caption{The physical setting of interests in three-dimensional space, in which (locally zooming in) the discrete charges are located at vertices of a equilateral triangular lattice.}
\label{fig:3D_Setting}
\end{figure}

In 3D, the electrical field $|\vec{E}|=q/4\pi \epsilon_0 r^2$ is created by a charged-particle of quantized charge $q$ at distance $r$, which is a consequence of the Gauss's law \cite{gauss1877theoria}. While the configuration with minimum electrostatic potential energy is not known in general, we can still try to estimate the depth $\lambda$ by assume the charged-particles are located at the vertices of an equilateral triangular lattice, locally, as $\mathcal{N} \rightarrow \infty$ \cite{bowick2002crystalline}. Under that assumption, the lattice spacing $a$ is roughly $a \approx \sqrt{16\pi / \sqrt{3} \mathcal{N}}$ and the electrical field at around the boundary for displacement $z$ in the radial direction from e.g. the midpoint of two neighbor vertices (see Fig. \ref{fig:3D_Setting}) can be approximated by:
\begin{equation}
E(1+z) \approx \frac12 + \frac1{\mathcal{N}a^3} \left[ \sum^{\infty}_n \frac{1}{\left(r_n/a \right)^3} \right] z + \mathcal{O}(z^2) \ ,
\label{E_in_z_3D}
\end{equation}
in which $r_n$ is the distance to the vertices of an infinite-flat equilateral triangular lattice. Note that now the unit-value of electrical field is $1/4\pi\epsilon_0=1$. 

\ \ 

\ \ 

For further analysis, we need to evaluate the infinite-series summation:
\begin{equation}
\mathcal{S} = \sum^{\infty}_n \frac{1}{\left(r_n/a \right)^3} \ .
\label{final_sum_3D}
\end{equation}
We can estimate the lower and upper bound of $\mathcal{S}$, which is done in Appendix \ref{app:sum_3D}. Numerical calculation \cite{matlab2021mathworks} using the vertices inside a square flat-area of size $10^4 a \times 10^4a$ gives $\mathcal{S} \approx 25.7$ in Eq. (\ref{E_in_z_3D}). We notice that this value is close to $8\pi$, off by about $2\%$. The depth $\lambda$ can therefore be estimated:
\begin{equation}
\lambda \sim \frac{E(1)}{\partial_z E(1+z) \Big|_{z=0}}  \approx \frac{a}{\sqrt{3}} \ .
\end{equation}
This is the same with the situation in 2D: the continuum limit breaks down when the discretization of charge becomes visible.

\ \ 

\section{Conclusion}

\ \ 

In summary, in this note we have shown evidences for the suppression of the electrical field inside a conductor as the number of charge-particles $\mathcal{N}$ goes up. There is a shallow region beneath the surface of the conductor in which the electrical field can still be viewed as non-zero, which depth $\lambda$ decreases monotonically with $\mathcal{N}$ (and in the same order of magnitude with the spacing between charged-particles). In usual experiment setting in introductory physics lab, a conducting spherical ball of radius $\sim 2$cm (the size of a ping-pong ball) with charge $\sim 1\mu$C (corresponds to $\sim 10^{13}$ electrons \cite{particle2020review}) will have the depth $\lambda \sim 10$nm, which is much larger than the typical inter-atomic spacing of metal and the Compton wavelength of an electron (which can be view as the quantum-smeared size of an electron) \cite{cohen19881986}!

\ \ 

\section{Acknowledgement}

\ \ 

We thank Tuan N. Truong and Minh D. Hoang for useful discussions. We also thank Duy V. Nguyen and the xPhO journal club for their support to share this finding to a wider audience. 

\ \ 

\appendix*

\section{Estimation of Eq. \eqref{final_sum_3D} \label{app:sum_3D}}

\ \ 

\begin{figure}[!htb]
\centering
\includegraphics[width=0.45\textwidth]{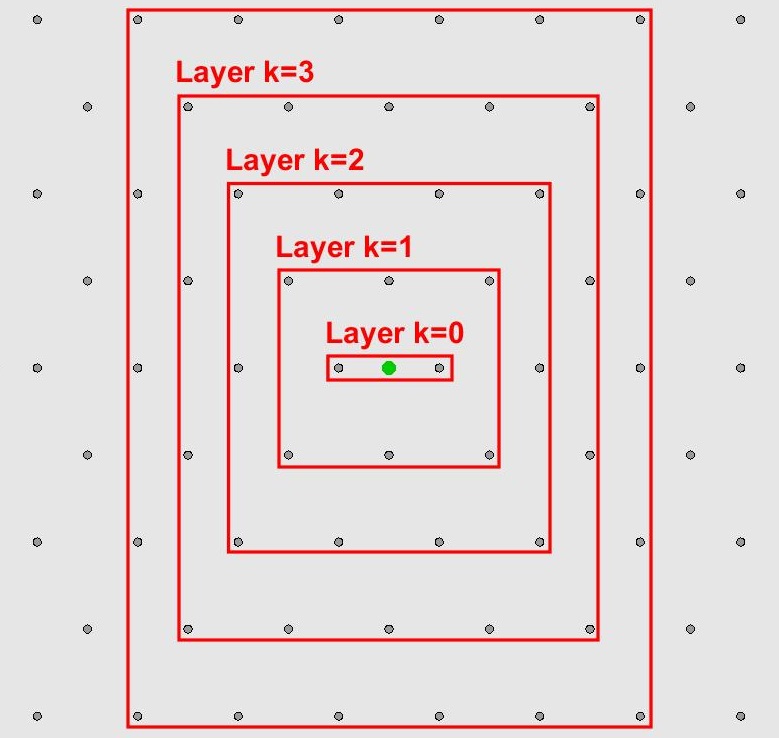}
\caption{The view (locally zooming in) near the surface from outside of the spherical conductor. We divide the charge-particles into layers as shown $k=0, 1, 2, ...$.}
\label{fig:3D_Setting_Up}
\end{figure}

Surrounding the point of interests on the surface of the spherical conductor, we can divide the charge-particles into layers $k=0, 1, 2, ...$ as shown in Fig. \ref{fig:3D_Setting_Up}. The number of charge-particles in the $k$-th layers is $C[k] =2(1+2k) $. For $k\geq 2$ the maximum and minimum distances between the point of interests and the $k$-th layer are $L_{max}[k]=k(\sqrt{3}a/2)$ and $L_{min}[k]=(k+1)(a/2)$. 

\ \ 

\ \ 

The contribution to $\mathcal{S}$ from the $0$-th and $1$-st layers is given by:
\begin{equation}
\begin{split}
\mathcal{S}_{o} &= 2\times\frac1{(1/2)^3} + 2\times\frac1{(\sqrt{3}/2)^3} + 4\times\frac1{(\sqrt{7}/2)^3}
\\
& = \frac{16}{441}\left( 441 + 49 \sqrt{3} + 18 \sqrt{7} \right) \approx 20.8 \ .
\end{split}
\end{equation}
We can thus obtain:
\begin{equation}
\sum^{\infty}_{k=2} \frac{C[k]}{\left(L_{max}[k]/a\right)^3} < \mathcal{S}-\mathcal{S}_{o} < \sum^{\infty}_{k=2} \frac{C[k]}{\left(L_{min}[k]/a \right)^3} \ .
\end{equation}
From the exact evaluations:
\begin{equation}
\sum^{\infty}_{k=2} \frac{C[k]}{\left(L_{max}[k]/a \right)^3} = \frac{16\left[\pi^2 - 9 + 3\zeta(3) \right]}{9\sqrt{3}} \approx 4.6 \ ,
\end{equation}
\begin{equation}
\sum^{\infty}_{k=2} \frac{C[k]}{\left(L_{min}[k]/a \right)^3} = \frac{16\pi^2}3 - 22 - 16\zeta(3) \approx 11.4 \ ,
\end{equation}
in which $\zeta(s)=\sum_{k=1}^{\infty} 1/k^{s}$ is the Euler-Riemann $\zeta$-function \cite{abramowitz1964handbook}, we have the lower and upper estimation:
\begin{equation}
25.4 < \mathcal{S} < 32.2 \ .
\end{equation}

\bibliography{main}
\bibliographystyle{apsrev4-2}

\end{document}